\pdfoutput=1
\documentclass[runningheads]{article}
\pdfoutput=1
\usepackage{fancyhdr} % To use headers
\usepackage{graphicx} %To use graphics package
\usepackage[textfont=bf, font = normalsize]{caption} %To use bold and normal size font for table and figure captions
\usepackage{floatrow} %To have table captions above the table
\usepackage{amssymb}
\usepackage{amsfonts}
\usepackage{amsmath}
\usepackage{multirow}
\usepackage{hyperref}
\usepackage{chemformula}
\usepackage{subfigure} % To use subfigures
\usepackage{todonotes}
\begin{document}

\title{Green HPC:\\
An analysis of the domain based on Top500}

\author{Abdessalam Benhari$^{1,2}$ \and Yves Denneulin$^{1}$ \and Frédéric Desprez$^{3}$\and Fanny Dufossé$^{1}$ \and Denis Trystram$^{1}$}

\date{%
    $^1$Univ. Grenoble Alpes, Grenoble INP, Inria, CNRS, LIG, France \\
    $^2$ Bull SAS (Eviden, Atos group), France \\
    $^3 $Inria, Grenoble, France \\[2ex]
    \today
}

\maketitle

\begin{abstract}
The demand in computing power has never stopped growing over the years. Today, the performance of the most powerful systems exceeds the exascale and the number of petascale systems continues to grow. 
Unfortunately, this growth also goes hand in hand with ever-increasing energy costs, which in turn means a significant carbon footprint. 
In view of the environmental crisis, this paper intents to look at the often hidden issue of energy consumption of HPC systems.
As it is not easy to access the data of the constructors, we then consider the Top500 as the \textit{tip of the iceberg} to identify the trends of the whole domain. 

The objective of this work is to analyze Top500 and Green500 data from several perspectives in order to identify the dynamic of the domain regarding its environmental impact. 
The contributions are to
take stock of the empirical laws governing the evolution of HPC computing systems both from the performance and energy perspectives,
to analyze the most relevant data for developing the performance and energy efficiency of large-scale computing systems,
to put these analyses into perspective with effects and impacts (lifespan of the HPC systems) and finally to derive a predictive model for the weight of HPC sector within the horizon 2030.

\textbf{keywords}: High Performance Computing, Energy efficiency , Top500, Green500
\end{abstract}

%%%%%%%%%%%%%%%%%%%%%%%%%%%%%%%%%%
\section{Introduction}
\label{sec:intro}

%\subsection{Brief presentation of the context}
%\label{subsec:context}

% contextualisation de la crise climatique et liens avec le domaine
With the climate crisis and the global warming, humanity faces an unprecedented challenge~\cite{IPCC2022}. 
It is also well-established that the ICT sector (Information and Communication Technologies) plays a significant role in this crisis through the huge energy consumption of computing, storage and interconnection devices. 
A recent survey of studies dealing with methodologies and tools for estimating the carbon footprint of the domain~\cite{Freitag2021} gives a value between 2.1 et 3.9\% of the global GHG emissions. 
In this wide variety of electronic and computing systems, data centers account for around one percent of the world's electricity consumption~\cite{masanet}. In France, the domain is estimated to reach 10\% by 2030~\cite{ademe}.
%\todo{Etude ADEME–Arcep sur l’empreinte environnementale du numérique en 2020, 2030 et 2050, ADEME, Mars 2023}

It is very difficult to have a precise view of the whole HPC domain. 
Many researchers in the field and political leaders believe that HPC will provide solutions to the warming/environmental crisis. 
Large-scale climate simulations are the only way to understand global dynamics and gain a clearer view of further decisions. 
Optimizing components in the fields of transportation, intelligent buildings, energy or health requires considerable computing power~\cite{rolnick}. 
On another hand, HPC is also part of the problem as it belongs to the digital world. 
In this paper, we develop an analysis that should contribute to understand the carbon footprint of the HPC domain.
Our main objective is to identify and clarify the impacts of HPC from the perspective of energy consumption. 
The main contributions are the following:
\begin{itemize}
    \item analyze the empirical laws governing the evolution of HPC computing systems both from the performance and energy perspectives,
    \item analyze the most relevant data to study the performance and energy efficiency of large-scale computing systems,
    \item put these analyses into perspective with effects and impacts (lifespan of HPC systems),
    \item derive an estimation of GHG emissions for the HPC domain within the horizon 2030.
\end{itemize}

%%%%%%%%%%%%%%%%%%%%%%%%%%%%%%%%%%%%%%%%%%%%%%%%%%%%%%%%%%%%%%%%%%%%%%%%
\section{Background}
\label{sec:background}
%Yves

Top500 was introduced in 1993 to rank the most powerful systems in the world using a common set of applications and criteria. It highlights the evolution and trends in HPC over the years. Including varying parameters on the architecture, the location and the capacity of systems, it has evolved to include the power consumption and the architecture's heterogeneity. It has rapidly gained in popularity and is now considered as a major source of information on the domain. Thoroughly studied over the years, many studies have observed the continuous evolution of computing capacity and energy efficiency on these systems. It was at the beginning fairly exhaustive but in the past 10 years has become less so because many powerful systems are not included, for example most hyperscalers ones. 
Nevertheless, the large duration and stability of the systems ranked in the Top500 makes it the most comprehensive and reliable source to study the evolution of high end powerful computers in the past 30 years. 
Over the years, new ranking lists have been released based on other benchmarks. Top500 introduced in 2017 a HPCG ranking based on a High Performance Conjugate Gradient benchmark, to bring more diversity to its evaluation. At the same epoch, Graph500 was released
with more data-intensive benchmarks. 
%There exist also dedicated benchmarks in specific domains like MLperf~\cite{MLperf} designed to the performance evaluation on Machine Learning. 
We focus in this study on Top500 data and we draft a comparison of evolution trajectory with Graph500. 

%In the following of this section, we introduce the main data-sets used in our analysis (the Top500 and Green500 data-sets). 
%Then, we open a discussion about the limitations of the study.

\subsection{Top500}
\label{subsec:top500}

\textit{Top500} is a project~\cite{TOP500} whose objective is to publish twice a year a ranking of the 500 most powerful HPC systems.
Throughout this time, it published more than fifty lists, in which we count over 10,000 supercomputers from more than 2,800 institutions around the world. 
%\todo{il faut rajouter que les listes sont revues systématiquement 2 fois pas an}
%Fanny: c'est dit dans la première phrase: twice a year
To rank supercomputers, the \textit{Top500 project} uses the well-known High-Performance \textit{Linpack}~\cite{Linpack} benchmark. A program supplied by Top500 has to be run on the HPC system. It solves a dense system of linear equations and provides the number of floating point operations per second (FLOPS) during its execution.

Top500 provides attributes about the identification and location of each system, its architecture and its performance. An HPC system is identified by its name, first and previous ranking, release year, and site.
Two attributes characterize each supercomputer's performance:
\textbf{$R_{max}$}, the maximal Linpack performance achieved
and \textbf{$R_{peak}$}, the theoretical peak performance calculated using the advertised clock rate of the CPU.
Concerning the architecture, it includes the number of cores, memory capacities, CPU/GPU types, constructors, etc. Table~\ref{table1} shows an example of Top500 main attributes for a recent HPC system. 
%However, the Top500 attributes list is not exhaustive. It lacks data on some architectural aspect of the presented supercomputers. 

\subsection{Green500}
\label{subsec:green500}

In 2008 energy efficiency became a concern in HPC. The monetary cost of running HPC systems exceeded the cost of purchasing and maintaining them~\cite{bp_cost}. This highlighted the importance of improving the energy efficiency of the upcoming supercomputers. 
 
To raise awareness on that matter, the \textit{Green500} was brought to light in 2009~\cite{Scogland2011}. This project started with three exploratory lists (Little Green500, HPCC Green500, and Open Green500), then became what we officially know as Green500. The official ranking published in the Top500 website starts from 2013 and is updated at the same pace as Top500. It uses the same Linpack benchmark as the {Top500 project}, but gives more attention to the energy efficiency metric to compare the different supercomputers displayed in the list. Other than the energy efficiency and the "green" ranking, Green500 contains the same attributes as Top500. 

\begin{table}[ht]
\caption{Example of Top500 (and Green500) main attributes.}%\vspace{-10pt}
%\resizebox{\linewidth}{!}{%
\begin{tabular}{|l|l|}
\hline 
\textbf{Attribute Name} & \textbf{Values} \\
\hline
Name & Frontier\\
Country & United States \\
Rank & 1 (year 2023)\\
Installation site & DOE/SC/Oak Ridge National Laboratory\\
Total cores & 8699904\\
Accelerator cores & 8138240\\
Processor  & AMD Optimized 3rd Generation EPYC 64C 2GHz\\
$R_{peak}$  & 1679818.75 $[TFlops]$\\
$R_{max}$ & 1194000 $[TFlops]$\\
Power  & 22703 $[kW]$\\
\hline
\end{tabular}%}
\label{table1}
\end{table}

\subsection{Graph500}
\label{subsec:graph500}

Graph500~\cite{graph500} is another ranking list of HPC systems based on more irregular computations related to graphs. 
It appeared in 2010 on BFS benchmark (for Breadth-First Search) including a GreenGraph500 variant. 
An extension list, namely SSSP (Single-Source Shortest Paths), was released in 2017. These benchmarks are data-intensive and not well suited for GPU execution. They are thus a counterpoint to the Linpack benchmark. The objective was to focus on real life applications with large-scale data-set from domains as cyber-security, medical data-base or symbolic network problems~\cite{graph500intro}. The metric of Graph500 is the Traversed Edges Per Second (TEPS).
Graph500 is however not as popular as Top500, and counts 231 systems in November 2023.
%Ces listes sont un pas vers les applis de la "vraie vie"
%On peut spécialiser aussi des benchmarks, comme ML-perf qui ont la volonté de trier les systèmes en vue des perf...
The focus of this paper is limited to the Top500. Graph500 will be only used for comparing the respective growth rates on different performance metrics.

\subsection{Restrictions}
\label{subsec:restrictions}

The main advantage of the Top500 is that it provides a stable view of the same features collected over many years and is so a valuable source to measure the evolution of HPC architectures.
Beside the fact that it only concerns a rather small part of the whole computing systems, it has two drawbacks:
\begin{description}
\item [its declarative character] the Top500 systems included in the list are submitted on a voluntary (not contractual) basis. 
The performances are self-valuated and self-reported with almost no independent checks. Many HPC systems are not included mainly for geopolitical or economical reasons.
Top500 does not contain all the most powerful systems,
in particular, some large companies do not have interest and thus, do not publish in the list.
\item [its orientation towards large scale HPC systems] this is especially true for the green counterpart of Top500 where highly innovative but small scale systems have no chance to appear.
\end{description}
%The main drawback is its declarative character. The TOP500 systems  included in the list are submitted on a voluntary (not contractual) basis. The performances are self-valuated and self-reported with almost no independent checks and many HPC systems are not in it mainly for political (China) or economical (companies in competitive industries) reasons.

%Peut-on dire ici que les Chinois en sont partis ? Autre chose, les grandes compagnies n'ont quasiment pas de réel intérêt à y contribuer...

%Another point is that they are oriented towards large scale HPC systems, this is especially true for the green counterpart of TOP500 where highly innovative but small scale systems have no chance to appear.

%One of the biggest "limitations" of the \textit{Top500} (and by transition the \textit{Green500}) data is its source: The \textit{Top500} relies on its standard protocol that is widely accepted, but self reported with almost no independent checks. This lowers the credibility of the data that these lists rely on.

%The other limitations are that the supercomputers ranked in the \textit{Green500} are the same as those in \textit{Top500} (with different rankings of course). This neglects all other efficient supercomputers outside the original list, which may result in some incoherent values in some graphs (Figure \ref{fig.1} for example).

\subsection{Evolution laws}

%Idée : d'avoir un outil simple qui prédit (grossièrement) l'évolution
%Moore, Koomey, lois empiriques...

What we refer to as a 'law' in this section is a trend observation done over many years that became widely adopted.
Such laws are used as a primary tool to provide a rough prediction about the evolution of a certain metric. 
The most popular ones are:

\textbf{Moore's law} states that the number of transistors in an integrated circuit doubles about every two years. 
It is also often extended to track the evolution of the performance of whole HPC systems \cite{Khan2021,Strohmaier2015,Subramaniam2013}.

\textbf{Koomey's law} is similar to Moore's law, but it targets the energy efficiency trend. 
It observes that the number of computations per joule of dissipated energy roughly doubles every 18 months. 
\cite{koomey}

%Projected over time, the evolution of these laws is exponential. 

%\todo{Parle-t-on de Dennard scaling (on l'utilise dans la suite...)? et qu'est-ce que l'on dit...}

%%%%%%%%%%%%%%%%%%%%%%%%%%%%%%%%%%%%%%%%%%%%%%%%%%%%%%%%%%%%%%%%%%%%%%%%
\section{Related works}
\label{sec:relatedworks}

%\todo{Deux volets : analyse de l'évolution, par des données du TOP500, principalement des mesures de perf, influence du hard (heterogene)

%Quelques articles proposent des projections de consommation énergétique, quel contour par rapport à ce que l'on va proposer ?}
%\todo{Un article parle d'ACV, lequel ?}

Some papers analyzed the performance of supercomputers based on Top500 and Green500. 
Many keynote talks also referred to the performance evolution of HPC systems.
However, none of them provide hints about the impact in term of carbon intensity, estimations and lifespan. 
%\todo{Il faut regarder le papier~\cite{projection2011}
%et comparer à aujourd'hui! : Projections jusqu'à 2024 par %classes d'architectures => les auteurs déduisent que seules %les architectures hétérogenes sont capables d'atteindre l %exascale (vrai pour frontier.. ) mais avec une energie qui %dépasse les 100 MW (faux: toujours le cas de frontier avec 21 %W}

\subsection{Trends / Performance analysis}

Each Top500 list is released every year in May during ISC High Performance conference and in November at SuperComputing.
It is the occasion to observe the evolution of trends in HPC. 
%Various studies gather results of these successive lists to observe the global evolution of the domain.
Regarding the technology, the authors of Top500 published a study in 2015~\cite{Strohmaier2015} using the evolution of $R_{max}$ over time to evaluate the increase of processors capacity compared to Moore's law. It describes key break points in the evolution which highlight a slowdown starting from 2008 (which is linked to growth rate of the number of components). 
This study is restricted to homogeneous systems and it was published before the era of GPUs. 
Milojicic et al.~\cite{Milojicic2021} studied the evolution trend of HPC from the architectural point of view. They showed that the HPC domain evolved to more customisation. It also highlights the performance slowdown mentioned in~\cite{Strohmaier2015} but attributes it to the end of Dennard scaling. 
Dennard scaling envisions a physical limit of power per die area that restrains the performance gain due to increasing density of transistors.
None of the above works really tackle the energy efficiency and its impact of the overall evolution of the HPC systems. 
Khan et al.~\cite{Khan2021} conducted an analysis that takes into account the architectural trends of the Top500. It focuses on the comparison between homogeneous and heterogeneous systems in term of performance and power consumption. It was mostly focused on the period 2009-2019, that includes the diffusion of GPU in HPC. In \cite{projection2011}, the authors made in 2011 performance and energy consumption projections up to 2024, claiming that only heterogeneous machines will be able to reach exascale but with an energy consumption up to 100 MW. It seems that Exascale was obtained with "only" 20 MW, but indeed with heterogeneous architectures. 

\subsection{Energy efficiency}

Green500 was introduced to
study the evolution of energetic efficiency of HPC systems, that is the ratio of $R_{max}$ over the total power consumption. 
In November 2023, the head of this list is Henri system, with more than 65 GFLOPS/watt.
In 2011, Scogland et al.~\cite{Scogland2011} described the evolution of the Green500 metric over the three first years to identify the design aspects that contribute to more power consumption in the objective of exascale. 
This study was extended in 2013 in~\cite{Subramaniam2013}. It introduces a new (holistic) metric to give a unified information about the energy efficiency/performance of Green500 called the EXASCALAR. 
%The authors also proposed a new evaluation metric for the Green500 list based on the HPL benchmark. Preciser un peu
This question of correlation between performance and energy efficiency of systems is crucial for various studies. 
Khan et al.~\cite{Khan2021} evaluated the evolution of the Pearson’s correlation
coefficient of the two lists over the period 2009-2019. 

Mair et al. \cite{Mair2015} analysed the energy efficiency of heterogeneous and homogeneous systems based on Green500, and proposed a new energy efficiency metric to avoid the bias induced by system sizes.
Hsu et al. mentioned the same metric in~\cite{Hsu2016}. 
They provide a 10 years retrospective on the evolution of energy efficiency metrics and their exploitation to evaluate HPC systems (mainly the Power Usage Efficiency and Performance-Power Ratio). 
It also highlights the main issues that need to be addressed in term of evaluation metrics and measurements.
Fraternali et al.~\cite{Fraternali2018} took a closer look at the impact of heterogeneity on performance and energy consumption variability by conducting directly their experiments on one of the top system in Green500. Although their study does not address trend analysis, they were able to propose some guidelines (hardware-wise) to build sustainable supercomputers.
Gao et al.~\cite{Gao2016} relayed on Top500 and Green500 to study the influence of design and architectural parameters on the Linpack scores in term of performance and energy efficiency. Then, they identified some development trends in supercomputer design.

%%%%%%%%%%%%%%%%%%%%%%%%%%%%%%%%%%%%%%%%%%%%%%%%%%%%%%%%%%%%%
\section{Methodology}
\label{sec:methodology}

%\subsection{Experimental protocol}
\label{subsec:protocol}

We present in this section the experimental protocol we followed for the results presented in this paper. 

%\todo{
%YD et DT : On ne comprend pas bien, merci d'expliciter}

\begin{description}
    \item[Dataset description] We collected data from three main existing lists (Top500, Green500 and Graph500). 
Then, they were filtered by removing any incomplete items or outliers (non-GPUs heterogeneous systems for instance). We also changed some of the resulting data to unify measurement units.  
Finally, the data were gathered into a single table to conduct the experiments where some new metrics were added (such as \textit{GFLOPS/watt, $R_{max}$/$R_{peak}$, etc.}).
    \item[Experimental setup] All the experiments were produced using a jupyter notebook\footnote{\url{https://jupyter.org/}} with a Python3.8 kernel and take around 30 seconds to reproduce.
    \item[Reproductibility] The code used to produce the different figures, along with all the steps to replicate these experiments is available on Github\footnote{\url{https://github.com/aBenhari/Green500-analysis.git}}.
 %   \item[Analysis \& Results] The collected data were analyzed focusing on two metrics, namely, the computing power ($R_{max}$ and $R_{peak}$) and the energy efficiency. Our goal was to investigate the usage trends and to provide short term projections. 
\end{description}

%ATTENTION : ceci est redondant avec les steps donnés plus bas ?... faut-il virer la subsection "steps" ?...

%TOP500 + Green comme source principale et on étent à d'autres benchmarks plus "compliqués" (plus irréguliers) Graph500
%Corpus de tests

%\subsection{Parameter choices}
%\label{subsec:law}
%C'est quoi le message ici ? 

%%%%%%%%%%%%%%%%%%%%%%

\section{Experiments}
\label{sec:experiments}

The following section reports the multiple experiments performed on the previously mentioned dataset to investigate the performance and energy efficiency trends of large-scale computing systems.

%The world of supercomputing is on a constant evolution in which certain trends start to manifest throughout the years. These trends point to the direction the world of supercomputing is going towards in terms of innovation, growth and progress for the next few years. The evolution of the energy efficiency and the performance are two of common trends that captured the interest of the community \cite{koomey,green500_source}. 

\subsection{Lifespan of systems in the Top500}
\label{subsec:lifespan}

%Qu'est-ce que l'on cherche à faire.
%We are interested in this experiments in à la longévité des systèmes dans le Top500

%Usage (conso en executant une appli) is not the unique source of energy.
%It is well-established that the material is significative while studying the energy footprint of a digital system
%\todo{DENIS ref : en gros 50\%}.

\begin{figure}[ht]
\centering 
\includegraphics[width=0.5\textwidth]{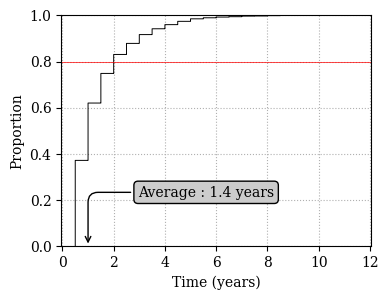}
\caption{lifespan of HPC systems in the Top500 (in years) from 1993. The red horizontal line means that 80\% of the systems stay less than 2 years.}
\label{fig:lifespan}
\end{figure}

We are interested here in the overall lifespan of HPC systems in Top500. 
We did an assessment of how long a system stays in Top500 after its first appearance by counting the number of apparitions of each system multiplied by 6, that corresponds to the interval between two successive publications of the list. 
%\todo{DENIS comment on retient/extrait la durée de vie des HPC depuis les listes. - On calcule le nombre d'apparitions de chaque machine, puis on multiplie cette valeur par 6 (c'est la durée entre chaque deux publications de listes)}
Figure~\ref{fig:lifespan} -- in which the \textit{x-axis} represents the apparition date in Top500, and the \textit{y-axis} represents the proportion of systems that stayed at (or more than) that time in Top500 -- summarizes the observed lifespan of these systems in Top500.

Only a selected few of the highest supercomputers on the list can appear in it for up to 8 years while $80 \%$ of them do not exceed an appearance time of $2$ years, which gives us an average of $1.4$ years. 
Although these results validate the rapid evolution of large-scale HPC systems, they do not reflect the real lifespan since we lose track of most of the systems after they leave the Top500. However, it provides an insight about the duration, which is an important metric to assess their ecological impact. Complete life cycle analyses show that the usage cost of an HPC system represents only a part of the total impacts~\cite{labos1.5}.
%\todo{impact for de la matérialité en gros 50\%}

Certain organizations choose to disclose information regarding their Supercomputer's architecture and usage. 
For instance, the cluster \textit{CURIE} displays its configuration with every upgrades or changes over time\footnote{\url{https://www-hpc.cea.fr/fr/Joliot-Curie.html}}.
While the exact lifespan of most HPC supercomputers remains uncertain, it typically spans between 5 to 7 years~\cite{labos1.5}. Thus, once they no longer figure in the list, these sytems are either upgraded to meet evolving computational demands (as is the case for \textit{CURIE}), decommissioned and disassembled, or repurposed for other uses. 

%\todo{A.B: developper ici une conclusion claire ! Que deviennent ces systèmes après être sortis du Top500, etc. }
 
%\todo{YD et DT : Re-introduire ici la Stability : Top1, Top 10, Top 100 ... avec un tableau} 

%%%%%%%%%%%%%%%%%%%%%%%%%%%%%%%
\subsection{Performance efficiency}
\label{subsec:perf}

\begin{figure}[ht]
\centering 
\includegraphics[width=1\textwidth]{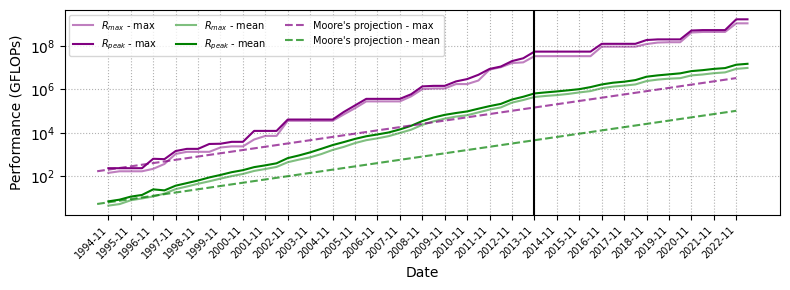}
\caption{ The evolution of the Top 1 [Purple lines] supercomputer's performance  metrics ($R_{max}, R_{peak}$) compared to the average evolution of all the Top500 supercomputer [Green lines]  by date, based on \textit{Top500} along with Moore’s law projection.}
\label{fig:evolutionTop500}
\end{figure}

In this experiment, our aim was to observe closely the performance evolution of the Top500 systems. We focused on the $R_{max}$ and $R_{peak}$ metrics for all the HPC systems on one hand, and the first system from each list on the other hand. 
Figure~\ref{fig:evolutionTop500} shows the evolution of both metrics in \textit{GFLOPS} (y-axis) over time (x-axis).

From this figure, a steady performance climb can be noticed through the years for the listed supercomputers, especially for those at the top. The $R_{max}$ and $R_{peak}$ have been increasing strongly since the beginning of the Top500 ranking. However, this exponential increase started slowing down since the second semester of 2013 where there is a clear breaking point.%, every period before and after follow a steady regression line.
We compared this growth to Moore's law projections. This comparison showed that overall, the performance evolution surpasses Moore's law with an average doubling time of $1.87$ years for the $R_{max}$ value instead of 2 years. However, this performance doubling is getting very close to Moore's projection after the previously mentioned break point.

\begin{figure}[ht]
\centering 
\includegraphics[width=1\textwidth]{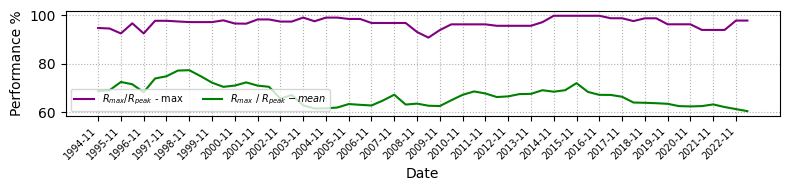}
\caption{ The performance ratio average between the Linpack $R_{max}$ and the theoretical $R_{peak}$ over time for the first \textit{Top500} system.}
\label{fig:rmax_sur_rpeak}
\end{figure} 

Another aspect we intended to highlight is the performance efficiency of the supercomputers listed in the Top500. For that, we calculated the ratio between the maximum performance reached using the Linpack benchmark ($R_{max}$) and the maximum theoretical performance ($R_{peak}$). The results are plotted in Figure~\ref{fig:rmax_sur_rpeak} to display the performance percentage ({y-axis}) evolution over time ({x-axis}). 
This graph shows a clear performance decrease in terms of percentage between the first Top500 supercomputer and the latter ones.
%\todo{Depuis 2015, on est en baisse (en cumul 10 \% !) et le plus mauvais resultat du Top500 , ceci veut dire que le logiciel est en "retard" sur le hard ???}
Other than energy or budget limitations, this decrease in growth and performance efficiency is correlated to the rise of architectural complexity of high performance computing systems that evolved significantly over the years in response to emerging applications requirements and technological improvements.

%On doit faire une sous-section sur les lois existantes :
%Ce sont des lois empiriques à partir d'observations...
%Moore historique (densité d'intégration des transistors), Moore étendue aux machines et aux plates-formes, Koomey, etc. 

%analysis: In both cases (maximum and average) the evolution of performance far exceeds Moore's Law, but we start observing that the evolution tends to approach it.

%%%%%%%%%%%%%%%%%%%%%%%%%%%%%%%%%%%%%%%%%%%
\subsection{Energy efficiency}
\label{subsec:energyefficiency}

Our purpose in this section is to highlight the progress made in terms of energy efficiency and to compare it with the performance.  We focused on Koomey's~\cite{koomey} law as a projection reference. 

%\todo{YD et DT : presenter ici la figure d'abord, et alléger la caption}

\begin{figure}[ht]
\centering 
\includegraphics[width=1\textwidth]{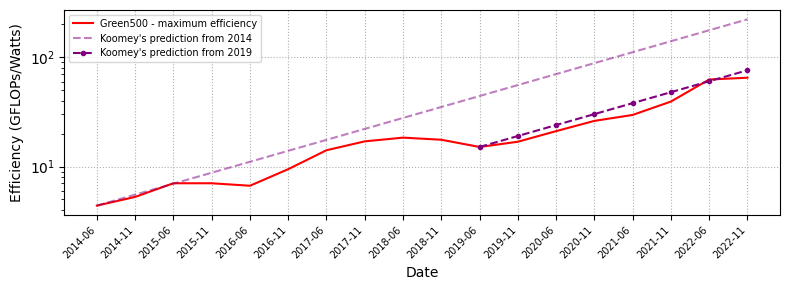}
\caption{Maximum \textit{Efficiency} of the \textit{Green500} 
supercomputers by list date along with Koomey's law projection starting at two different periods [2014 and 2019].}
\label{fig:Koomey}
% \label{fig.fig.3}
\end{figure}

{Figure}\ref{fig:Koomey} illustrates a first experiment where we compare the  energy efficiency evolution of HPC supercomputers in GFLOPS/Watt (\textit{y-axis}) to Koomey's law projections from two starting points (2014 and 2019) over time (\textit{x-axis}).

We observe an increase in the energy efficiency, but at a lower rate than the performance. 
A huge progress have been made in that matter, with a maximum value increase from $4.5$ $GFLOPS/watt$ in the late 2013 to $65.39$ $GFLOPS/watt$ in 2023. However, this progress fails to follow Koomey's law with a doubling time higher than 1.5 years, especially in the beginning of {Green500}. Fortunately, it seems to catch up in the last few years due to rising interest for energy efficiency in the HPC community and the emergence of heterogeneous supercomputers that appear to be more energy efficient.

\begin{figure}[ht]
\centering 
\includegraphics[width=\textwidth]{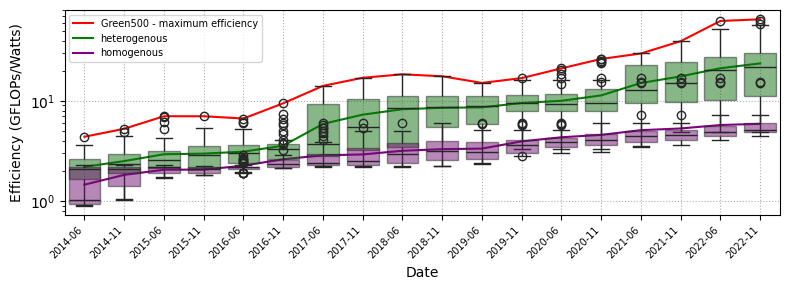}
\caption{Maximum Energy efficiency growth \textit{Green500} systems by date distinguished by architecture type (homogeneous vs heterogeneous)}
\label{fig:heterogeneity}
\end{figure} 

To take a closer look on the impact of heterogeneity on the energy efficiency of Green500, we propose a second experiment where we compare the heterogeneous and homogeneous systems in that same list. Our results are displayed in Figure~\ref{fig:heterogeneity} where the {y-axis} represents the energy efficiency and the {x-axis} represents time. This figure definitely shows that homogeneous systems are less efficient than heterogeneous ones by an order of magnitude and that the gap between the two categories is increasing. A lot of architectural work has been done to develop energy efficient chips for vector computations.
It confirms that using dedicated architectures for certain operations, typically vectorial ones, is a good way to increase efficiency. This is also a trend in lower end systems like smartphones with SoCs including components for data analysis and tensoral computations capabilities. Nonetheless heterogeneous systems are harder to program efficiently than homogeneous ones. Exploiting them to their full potential thus requires significant human expertise that all sites can not afford. We think that this explains the recent decrease of efficiency on average in the Top500 while the top system remains at a very high level as shown in Figure~\ref{fig:rmax_sur_rpeak}.
%\todo{paragraphe un peu spéculatif et polémique, je suis preneur de vos avis -- Yves}

%comparaison entre l'evolution des machines heterogenes et homogenes du Green500 en termes d'efficacité energétique.
        
%analyse ? En voit clairement une dégradation de l'efficacite energetique des machines HPC homogenes par rapport aux machines heterogenes au cours du temps, leur archi (heterogenes) leur permet d'etre plus efficace

%\todo{YD : detailler rapidment l'analyse finale}

%%%%%%%%%%%%%%%%%%%%%%%%%%%%%%%%%%%%%%%%%%%%%%%%%%%%
\subsection{Does Top500 really represent real life?}
\label{subsec:variants}

%Other benchmarks: Graph500 and variants
%\todo{Un mot d'intro, TOP500, évalué sur son benchmark Linpack est très régulier, on investigue ici des bench plus généraux...}

The evolution of $R_{max}$ performance on Top500 is mostly driven by technology improvements, as highlighted in previous figures. 
One can expect the same evolution on other benchmarks, including less regular ones. 
Table~\ref{Tab:list-ratio} presents the improvement ratio over 12 years for various benchmarks. 
We consider the ranking lists corresponding to the two benchmarks of Top500: Linpack ($R_{max}$) and HPCG, and two benchmarks of Graph500: BFS and SSSP. 
For each list, we indicate the increase ratio over 3 periods of 4 years (11/2010-11/2014, 11/2014-11/2018 and 11/2018-11/2022) for the best system of the corresponding list (top 1), for the 5, 15 first systems. 
As HPCG list and SSSP list appeared in 2017, they are only evaluated over the second period. 
The list is extended to the 500 systems for only Top500 as Graph500 does not list 500 systems. BFS increase ratio is not given for the first period, as it only contained 10 systems in 2010, and the increase ratio was over 3,000 over the period for the first systems.

\begin{table}[ht]
        \begin{tabular}{|l|l|l|l|l|l|}
                \hline
                &&&2010-2014&2014-2018&2018-2022\\
                \hline
                \multirow{8}{*}{Top500}&\multirow{4}{*}{$R_{max}$ (FLOPS)}&top 1&13.20&4.24&7.68\\
                \cline{3-6}
                &&top 5&11.19&4.72&5.26\\
                \cline{3-6}
                &&top 15&8.50&4.53&4.75\\
                \cline{3-6}
                &&top 500&7.07&4.58&3.44\\
                \cline{2-6}
                &\multirow{4}{*}{HPCG (FLOPS)}&top 1&-&-&5.47\\
                \cline{3-6}
                &&top 5&-&-&6.11\\
                \cline{3-6}
                &&top 15&-&-&5.03\\
                \cline{3-6}
                &&top 500&-&-&4.99\\
                \hline
                \hline
                \multirow{6}{*}{Graph500}&&top 1&-&1.32&3.29\\
                \cline{3-6}
                &BFS (TEPS)&top 5&-&1.50&1.79\\
                \cline{3-6}
                &&top 15&-&1.57&1.76\\
                \cline{2-6}
                &&top 1&-&-&11.81\\
                \cline{3-6}
                &SSSP (TEPS)&top 5&-&-&54.18\\
                \cline{3-6}
                &&top 15&-&-&186.98\\
                \hline
        \end{tabular}
        \caption{Ratio of increase over 4 years periods per list between 2010 and 2022  in Top500 and Graph500.}
        \label{Tab:list-ratio}
\end{table}

A large difference can be observed between the different benchmarks. On the period 2018-2022, the increase ratio for vary between 1.76 for the top 15 of BFS benchmark and 186.98 for the top 15 of SSSP benchmark. It could be explained by the novelty of SSSP in 2018 and a wrong parametrization in the first years. However, BFS and $R_{max}$ also behave differently. Top 1 $R_{max}$ performance increased around 7.5 times during the same period, while BFS increased around 3.3 times only. BFS benchmark is more data-intensive than Linpack. The slowdown of this list compared to $R_{max}$ could be the result of this point.
On the contrary, for each list and period, we observe quite similar ratio for top 1, top 5, and top 15. If the best performance globally grows faster than the mean performance as observed in Figure~\ref{fig:evolutionTop500} it is not the case for period 2014-2018 and for HPCG.
Finally, a slowdown was observed after 2014. This focus on period 2014-2022 shows an acceleration on top systems after 2018. The comparison with Figure~\ref{fig:Koomey} highlights the impact of accelerators arrival in HPC.

%Parle-t-on de la variante TOP500 HPCG ?

%%%%%%%%%%%%%%%%%%%%%%%%%%%%%%%%%%%%%%%%%%%%%%%%%%%%%
\section{Projection model}
\label{sec:projection}

\subsection{Footprint efficiency}
\label{subsec:footprint}

The carbon footprint of HPC electricity consumption was the main objective behind Green500 release. It is however mostly studied for energy efficiency. We consider in this section the carbon footprint corresponding to Green500 power values. We compute the footprint of an HPC system by multiplying its power consumption by the electricity footprint of its country at the corresponding year\footnote{\url{https://ourworldindata.org/grapher/carbon-intensity-electricity}}.
\begin{figure}[ht]
\centering 
\includegraphics[width=1\textwidth]{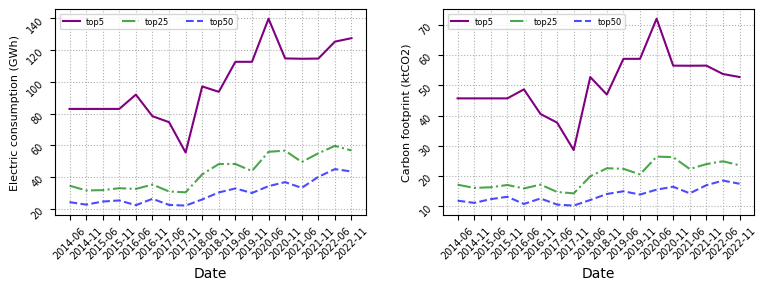}
\caption{Electric consumption \& carbon footprint for the top 5, 25, 50 in Top500 over time.}
\label{top500_conso_co2}
\end{figure}
%\todo{convertir la deuxième partie de la figure 6 en kilotonnes (kt) => Done}

Figure~\ref{top500_conso_co2} shows the evolution of mean electricity consumption of the first Top500 systems since 2014 and the corresponding mean footprint. It highlights the big difference of the top 5 systems and the following ones. The consumption was almost constant before 2018 and was multiplied by 2 until 2022. The increase is restrained for the footprint due to a reduction of electricity footprint in many countries at the same period.
Both curves are very close except for the very recent period due to the energy mix. %(Frontier is located in a state with hydroelectric dams and nuclear power plants). 
%Fanny: on calcule par pays et pas par état... ou alors on parle de LUMI en Finlande, Top 5 avec moins de 150g/kwh dans le pays...
\begin{figure}[ht]
\centering 
\includegraphics[width=1\textwidth]{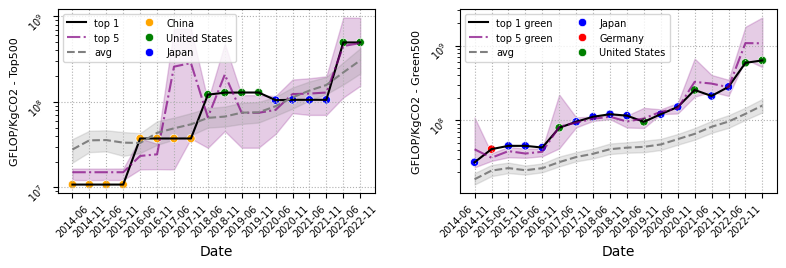}
\caption{Evolution of the maximum performance in GFLOP per kg\ch{CO2} for the top 1 \& 5  systems in the Top500 [Left] and Green500 [right].}
\label{gflops_per_kgco2}
\end{figure} 

Figure~\ref{gflops_per_kgco2} shows the evolution of carbon efficiency in Top500, that is the number of FLOP per kg of \ch{CO2}. 
We observe a regular increase of the mean efficiency since 2014 with doubling period of 2.83 years on average.
%\todo{Abdessalam, tu peux calculer la valeur correcte et modifier? => Oui ! :  c'est environ 2.83 ans}
The first Top500 systems are close to the mean while the efficiency of the Green500 increases faster.
%\todo{Fanny: C'est quoi ta remarque ? Abdessalam?}

%\todo[inline]{on extrait du Top500 les consos electriques directement du tableau et le pays, on en déduit les émissions (à partir du mix energetic par pays et par année)}

%\begin{figure}[ht]
%\centering 
%\includegraphics[width=1\textwidth]{./green500_conso_co2.png}
%\caption{Electric consumption \& carbon footprint for the top 1 in Green500 over time (highlighted by country).}
%\label{green500_conso_co2}
%\end{figure}

%%%%%%%%%%%%%%%%%%%%%%%%%
\subsection{Scenarios}
\label{subsec:scenqarios}

Based on the previous analysis, we provide an estimation of the electricity consumption of the domain in order to compare with the roadmap for decarbonization.
Various scenarios exist, for instance those provided by IPCC\cite{IPCC2022}.
The objective of the European Union Green Deal~\cite{europeancommissionDeliveringEuropeanGreen2021} is to reduce the GHG emissions by 55\% by 2030. 
We target the 2030 horizon since it is unreasonable to look any further. 
By extrapolating from the carbon efficiency data displayed in Fig.~\ref{gflops_per_kgco2}, we estimate the increase of the number of GFLOP per kg\ch{CO2} to reach $1.64 * 10^9$ in 2030 compared to $3.05 * 10^8$ in 2022, representing an increase of 537\% between 2022 and 2030 with a slope of 11.87\% per issue date, which corresponds to 24.99\% per year of the value recorded in 2022. 
Of course, this is a rough estimate, but our results have shown great stability of this improvement over a long period of time. 
The gap between this increase and the expected reduction is huge. 

%\todo[inline]{On se base d'abord sur les data existantes pour le domaine ICT entier, et aussi de notre analyse précédente, on en déduit une évolution par scénarios (Freitag Patterns 2021). 
%
%TRAVAIL A FAIRE :
%courbe d'évolution du Top500 à partir de l'historique (même tendance depuis très longtemps)
%en émissions Carbone.
%On conclue sur un roadmap/ Comparer aux scénarios IPCC de 2030, et on conclut sur une discussion comment y arriver (externalités positives).
%AB calcule le HPC en 2030 et la pente (pourcentage de 2023 ou 2022) : Projection (très optimiste) sur 2030 = $1.64 * 10^9$ comparée à $3.05 * 10^8$ soit une augmentation de \textbf{537 \%} entre 2022 et 2023 avec une pente de\textbf{ 29.19 \% }de la valeur de 2022
%
%on donne l'objectif de IPCC en 2030 en terme de decrease.
%On compare}

% Dans un second temps, pour la projection, on part du point de référence en 2030 et on montre la trajectoire qu'il nous faudrait pour y arriver...

%%%%%%%%%%%%%%%%%%%%%%%%%%%%%%%%%%%%%%%%%%%%%%%%%%%%%
\section{Conclusion}
\label{sec:conclusion}

In this paper, we have studied the evolution of the HPC domain through data from the Top500 and its variants
which we consider to be representative of the field. 
In-depth analysis of both performance and energy efficiency has enabled us to better understand the evolution
and highlights several important results:
(i) the lifespan of HPC systems in Top500 is in average lower than 2 years and has not evolved in the past decades, (ii) the performance and energy efficiency increase has diminished between 2014 and 2018, but keep increasing and (iii) heterogeneous systems are potentially more energy efficient but require more human expertise to be fully exploited.
%\todo{muscler les "important results"}

%Dans cet article, nous avons étudié l'évolution du secteur HPC à travers les données du Top500 et ses variantes
%que l'on considère comme représentatif du domaine. 
%L'analyse approfondie sur les deux plans de la performance et de l'efficacité énergétique a permis de comprendre les points importants.  (reprendre la liste des sous-sections 5)

In addition, we have proposed a prospective study up to the 2030. 
Consumption of HPC systems is set to continue rising, as it has been doing for a long time, at a time when GHG emissions need to be drastically reduced and the gap is huge. 
However, the calculations carried out on these large-scale systems may also help us to propose solutions to mitigate the crisis. 
Nowadays IPCC experts consider that it is still time to act against climate change. 
HPC domain needs to participate to the effort by improving its footprint. 
We hope this study will help estimating the necessary efforts to meet IPCC roadmaps.

%Climate-positive HPC applications should be encouraged and the other applications should be avoided to have a chance to achieve a positive overall balance.
%En plus, on propose une étude prospective pour se projeter dans les années à venir. 
%Message positif : il est encore temps : il faudrait que les applis HPC positives pour mitiger le CO2 soient très conséquentes.
%L'effort à faire n'est peut être pas si grand car au final, notre étude a montré que la consommation électrique n'a pas beaucoup augmentée.

%next step ? Il faudrait aller sur des analyses de comportement et du rebond, mais c'est très dur...

%\todo{pistes pour réduire la longueur :

%- virer la phrase du début sur MLperf + la ref

%- virer les acknowledgements }

\section{Acknowledgments}
Abdessalam Benhari is supported by the ANRT CIFRE convention 2022/0176.
This work was supported by the research program on Edge Intelligence of the Multi-disciplinary Institute on Artificial Intelligence MIAI at Grenoble Alpes (ANR-19-P3IA-0003) and by the REGALE (H2020-JTI-EuroHPC-2019-1 agreement n. 956560) european project. 

\bibliographystyle{splncs04}
\bibliography{biblio.bib}

\end{document}